# Observing the completion of the first solvation shell of carbon dioxide in argon from rotationally resolved spectra

A.J. Barclay,[1] A.R.W. McKellar,[2] and N. Moazzen-Ahmadi[1]

[1] *Department of Physics and Astronomy, University of Calgary, 2500 University Drive North West, Calgary, Alberta T2N 1N4, Canada*

[2]*National Research Council of Canada, Ottawa, Ontario K1A 0R6, Canada*

**Abstract**

Widespread interest in weakly bound molecular clusters of medium size (5 ~ 50 molecules) is motivated by their complicated energy landscapes, which lead to hundreds or thousands of distinct isomers. But most studies are theoretical in nature, and there are no experimental results which provide definitive structural information on completion of the first solvation shell. Here we assign rotationally resolved mid-infrared spectra to argon clusters containing a single carbon dioxide molecule, $CO_2$-$Ar_{15}$ and $CO_2$-$Ar_{17}$. These mark completion of the first solvation shell for $CO_2$ in argon. The assignments are confirmed by nuclear spin intensity alternation in the spectra, a marker of highly symmetric structures for these clusters. Precise values are determined for rotational parameters, and for shifts of the $CO_2$ vibrational frequency induced by the argon atoms. The spectra indicate possible low frequency (≈2 cm$^{-1}$) vibrational modes in these clusters, posing a challenge for future cluster theory.

There has long been interest in the structures and energetics of weakly bound molecular clusters. Even relatively small clusters governed by simple forces, such as a Lennard Jones potential, have complicated energy landscapes involving many isomers (local minima).[1] Detailed understanding of such phenomena at the microscopic level is essential for manipulation of chemical reactions, modulating biological activities and solvation effects. In the particular case of the $CO_2$-$Ar_n$ system, there have been at least six publications concerned with theoretical structures,[2-7] but there are no experimental results for clusters beyond n = 2. Direct high-level *ab initio* calculations become very challenging for larger clusters (many electrons!). The use of two-body intermolecular potentials is more practical, but neglects possible non-additive effects on energies and geometries. As well, the number of local minima on the potential hypersurface rises exponentially with cluster size, so that there are hundreds or even thousands of distinct isomers for medium size clusters like those considered here. One challenge is to locate the most stable global minimum with some confidence. Experimental results, which usually detect the most stable isomer, are clearly helpful for validation of theoretical methods. Experimentally, however, the challenges are to generate clusters consisting of 10 or more monomers at high abundance and obtain spectral signatures at rotational resolution for definitive structural identification.

Here we report high resolution infrared spectra of $CO_2$-$Ar_{15}$ and $CO_2$-$Ar_{17}$, the latter marking the completion of the first solvation shell of Ar around the $CO_2$ molecule. For each cluster, we obtain precise information on the shift of the $CO_2$ $\nu_3$ (asymmetric stretch) vibrational mode induced by the surrounding Ar atoms, and on the $B$ rotational constant, which basically relates to the physical size of the cluster. As will be shown, the nature of the observed spectra (with nuclear spin intensity alternation) confirms highly symmetric geometries for each cluster.

$CO_2$-Ar was originally studied in the microwave region.[8] A T-shaped structure was established, with the Ar atom located to the ‘side’ of the $CO_2$ molecule, adjacent to the C atom at an intermolecular separation of about 3.5 Å. Since then, there have been many further microwave[9,10] and infrared[11-17] studies. The trimer $CO_2$-$Ar_2$ has also been detected in the microwave[18] and infrared[2] regions. Its structure locates the second Ar atom in a position equivalent to the first, giving a $C_{2v}$ point group structure with an Ar-Ar distance of about 3.8 Å. There are numerous theoretical investigations of the $CO_2$-Ar interaction potential.[19-25]

Experimental information on doped rare gas clusters with n > 3 is in short supply with the notable exception of helium, whose superfluid nature has allowed spectroscopy of a whole range of cluster sizes and dopant molecules. Thus for $CO_2$-$(He)_n$, spectra have been observed for n = 1 to 60.[26,27] Another exception is HF-$Ar_n$, where gas phase spectra have been observed for n = 1 - 4,[28,29] and helium nanodroplet spectra for n =1- 6 and beyond. In the latter work of Nauta and Miller[30] rotational resolution was not possible in the nanodroplet environment, and, as the authors noted: "… the resulting [cluster] structures can be quite different from those obtained from gas phase nucleation".[30] In any case, the large dipole moment of HF makes it a rather different probe than $CO_2$. Theoretical studies of HF-$Ar_n$ clusters can be found in Refs. [31-33].

Our spectra were recorded as described previously,[17,34] using a pulsed supersonic slit jet expansion probed by a rapid-scan optical parametric oscillator source. The slit width was set at 12.5 μm, narrower than we have previously used in order to encourage the formation of larger clusters. The combination of the narrower slit width, the use of a multichannel block assembly, careful adjustment of the backing pressure, and the use of quantum-correlated twin beams (idler and signal) for cancellation of the power fluctuations[35] made it possible to record spectra at high signal to noise for the clusters discussed here. The gas expansion mixture contained about 0.03% carbon dioxide plus 1% argon in helium carrier gas with a backing pressure of about 20 atmospheres. Wavenumber calibration was carried out by simultaneously recording signals from a fixed etalon and a reference gas cell containing room temperature $CO_2$. Spectral simulation and fitting were made using the PGOPHER software.[36] We carried out our own cluster calculations[34] since those in the literature[2-6] often do not include full details on bond lengths, rotational constants, and binding energies. For these, we used the Ar-Ar interaction potential of Dieters and Sadus,[37] together with the *ab initio* $CO_2$-Ar potential energy surfaces for the $CO_2$ ground and $v_3$ = 1 states as calculated by Cui et al. [23] Since the same $CO_2$-Ar potential was used by Wang and Xie for their $CO_2$-$Ar_n$ calculations,[6,7] our cluster results are similar to theirs (with small differences in calculated binding energies due to the use of different Ar-Ar potentials).

We have obtained experimental infrared spectra for a range of $CO_2$-$Ar_n$ cluster sizes, and here we focus on the largest ones for which clear assignments are presently possible, n = 15 and 17. Our calculated global minimum structures for the most stable isomers are shown in Figs. 1 and 2 and the calculated rotational constants are listed in Table 1. We find both clusters to have highly symmetric cage-like structures with $D_{3h}$ and $D_{5h}$ point group symmetry, respectively. $CO_2$-$Ar_{15}$ has five argon rings around the $CO_2$ axis, each consisting of three Ar atoms. This agrees with Wang and Xie[6,7] and probably with Severson[3] and Böyükata et al.,[4] as well, though it is difficult to be sure from their figures. $CO_2$-$Ar_{17}$ has three concentric rings, each consisting of five Ar atoms, plus two additional Ar atoms, one located at each end. This clearly agrees with all the previous studies. Structures from Jose and Gadre[5] are not available because the largest cluster in their study is $CO_2$-$Ar_{12}$. Our results give *equilibrium* structures, and we know that the real structures will have slightly longer effective bond lengths as a result of zero-point motions.

Table 1. Molecular parameters for $CO_2$-$Ar_{15}$and $CO_2$-$Ar_{17}$ (in $cm^{-1}$).[a]

| | $CO_2$-$Ar_{15}$ | | $CO_2$-$Ar_{17}$ | |
|---|---|---|---|---|
| | Experiment | Theory (present) | Experiment | Theory (present) |
| Origin | 2341.9804(1) | | 2340.4719(1) | |
| $A$ | [0.00305] | 0.00314 | [0.00260] | 0.00270 |
| $B''$ | 0.0023325(8) | 0.00240 | 0.0018185(3) | 0.00189 |
| $B' - B''$ | $-1.09(4)\times10^{-6}$ | | [0.0] | |
| $D_J$ | $1.8(4)\times10^{-9}$ | | [0.0] | |

[a] Quantities in parentheses are 1σ from the least-squares fits, in units of the last quoted digit. Quantities in square brackets were fixed in the fits.

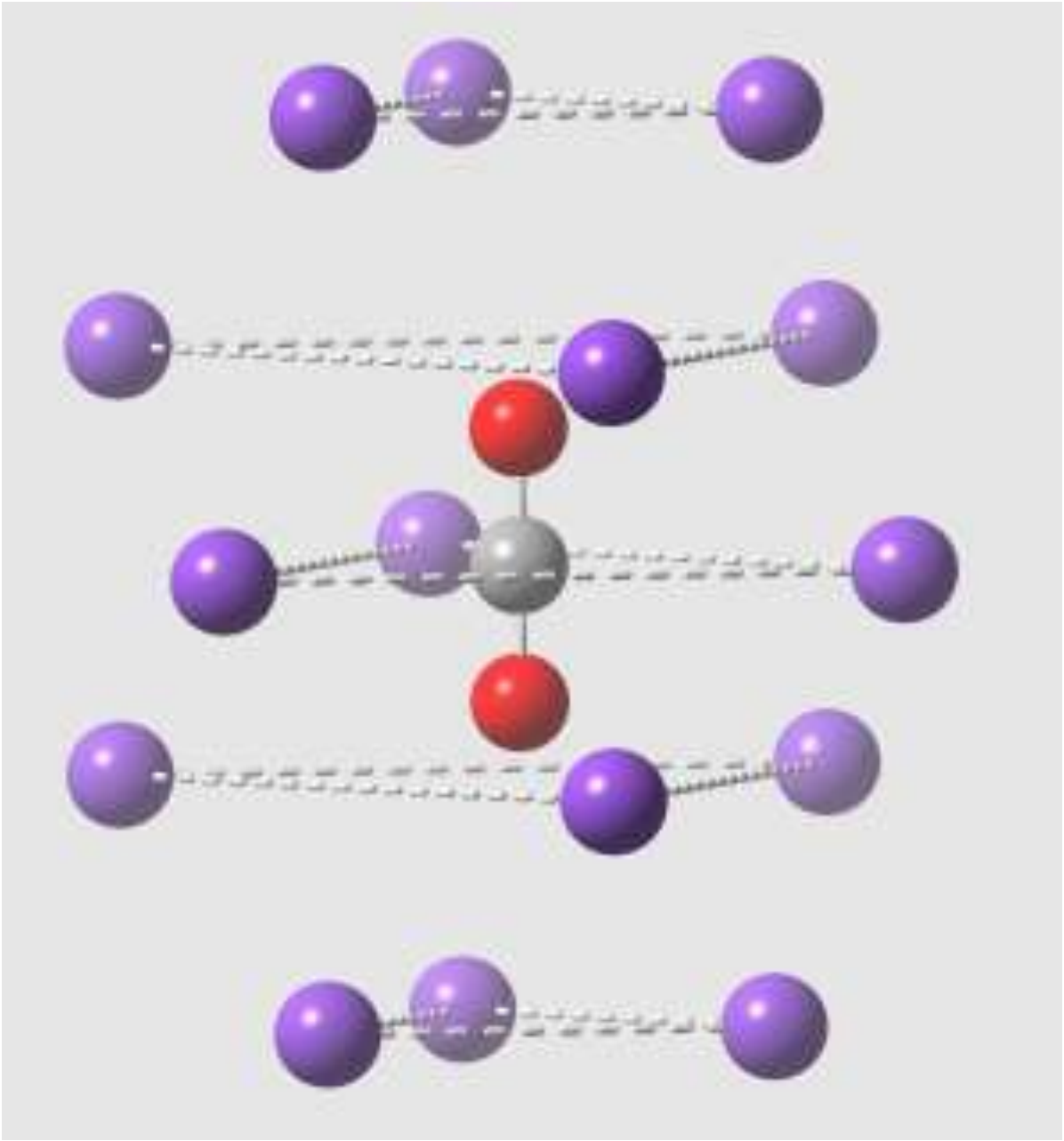

Fig. 1: The calculated global minimum equilibrium structure of $CO_2$-$Ar_{15}$.

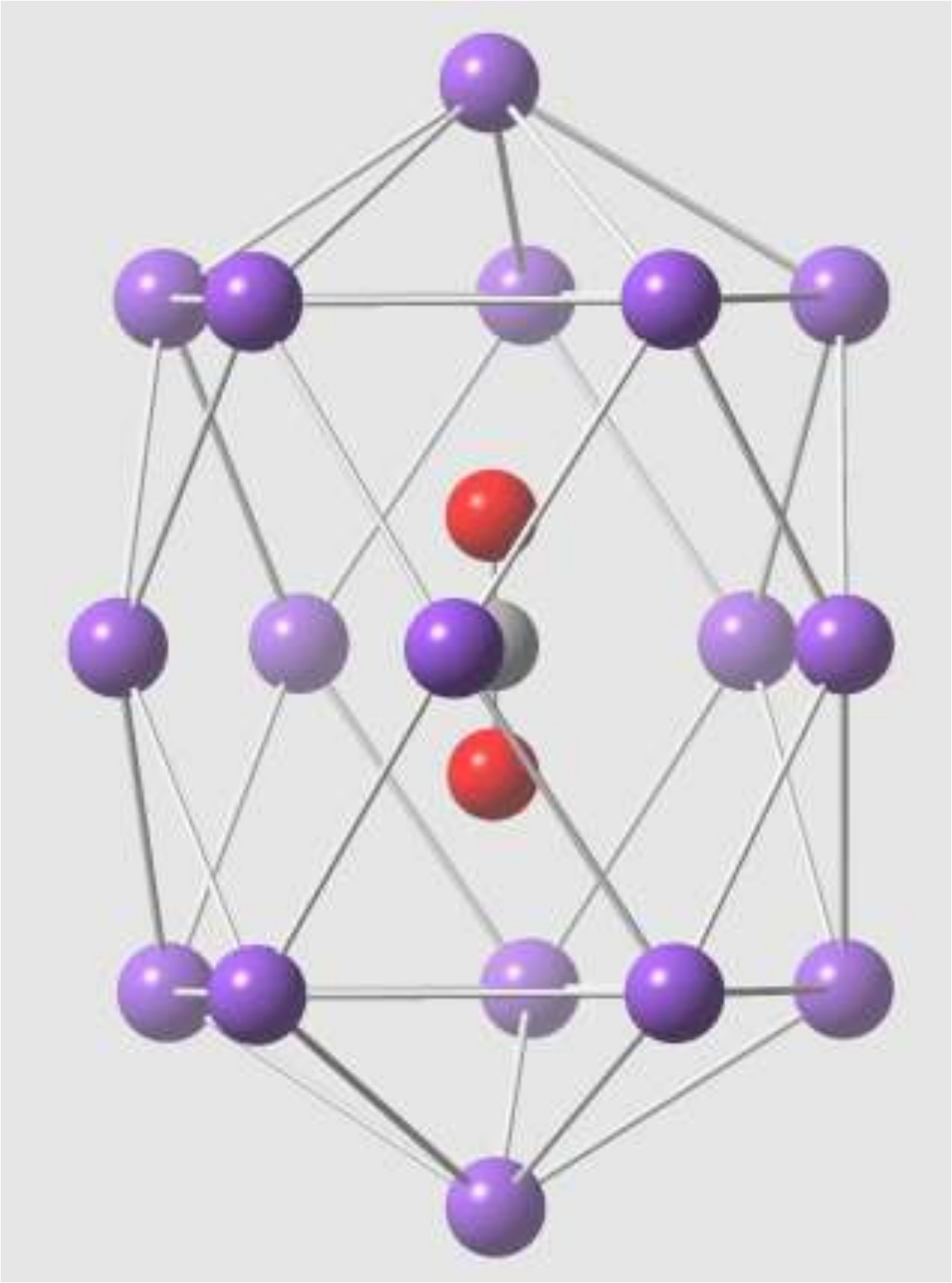

Fig. 2: The calculated global minimum equilibrium structure of $CO_2$-$Ar_{17}$.

The observed spectral bands that we assign to $CO_2$-$Ar_{15}$ and $CO_2$-$Ar_{17}$ are shown in Figs. 3 and 4, respectively, and parameters resulting from their analysis are listed in Table 1. The experimental rotational constants *B* are quite well determined, and have values about 3% smaller than our calculated equilibrium constants. This difference agrees well with that expected between equilibrium and zero-point parameters: for example, experimental rotational constants[2,18] of $CO_2$-$Ar_2$ are 2.6 to 4.2% smaller than those from a similar equilibrium calculation using the same potential functions. For symmetric rotor molecules, the spectra are not sensitive to the values of the *A* rotational constant, so when analyzing them we used calculated *A* values scaled by the same factors as observed for *B*.

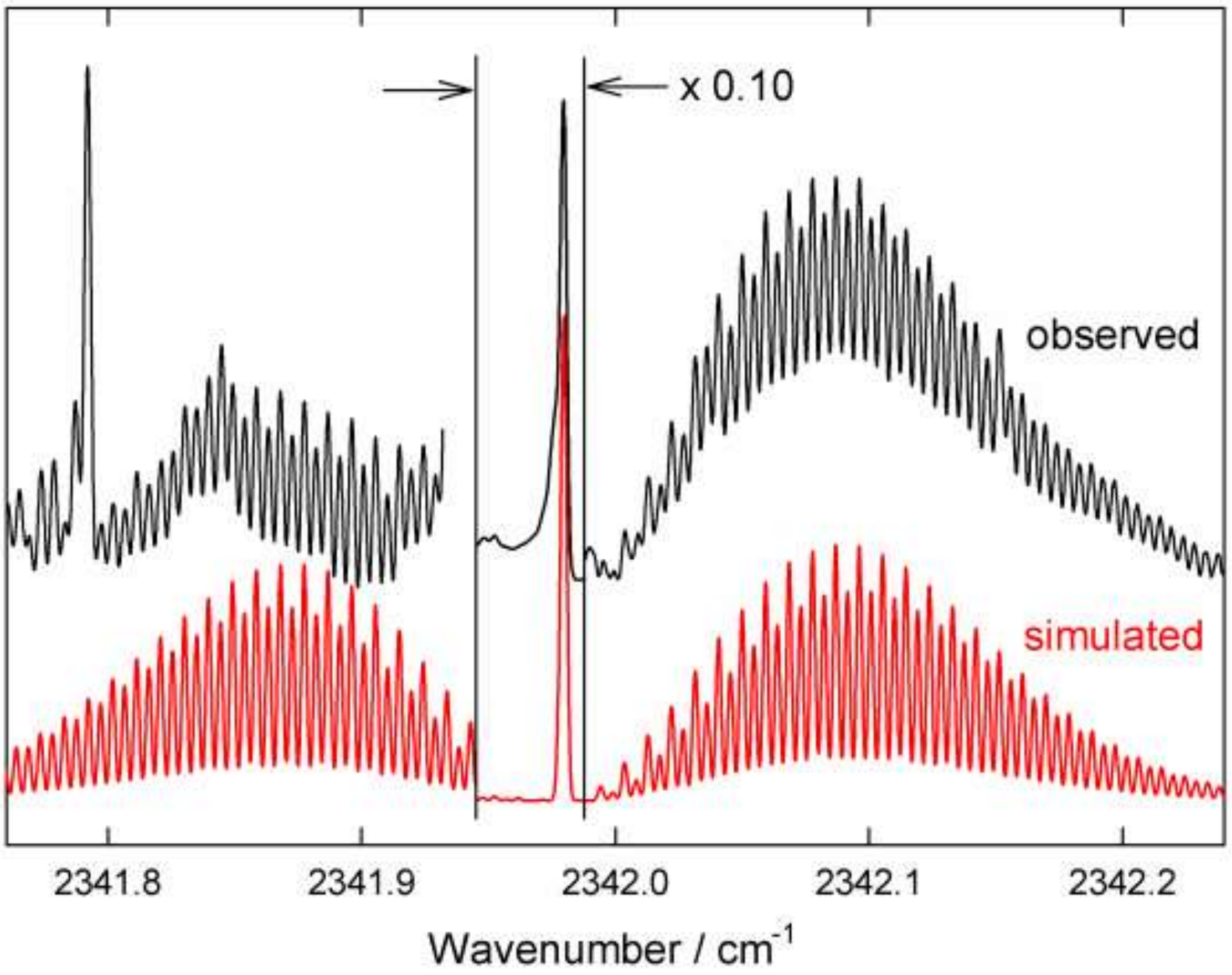

Figure 3: Observed and simulated (T = 2.0 K) spectra assigned to $CO_2$-$Ar_{15}$. The gap in the observed spectrum is due to a $CO_2$ monomer line (*R*(6) of the ($01^11$)-($01^10$) hot band). The unidentified line at 2341.792 $cm^{-1}$ is probably due to a different size $CO_2$-$Ar_n$ cluster. The low frequency shoulder on the central *Q*-branch in the observed spectrum may be analogous to the *Q*-branch series observed for $CO_2$-$Ar_{17}$ which we attribute to sequence bands (see text).

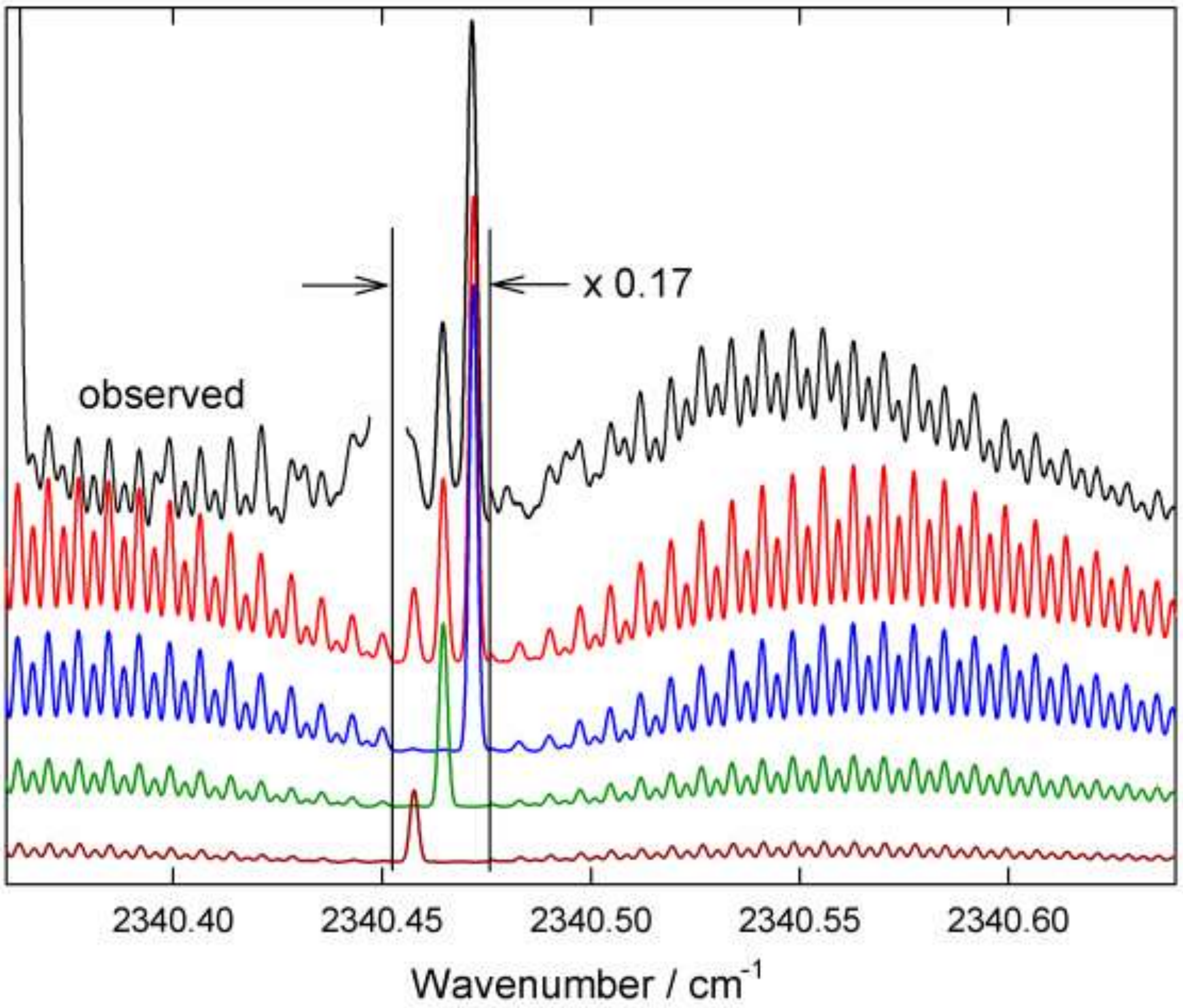

Figure 4: Observed and simulated (T = 2.0 K) spectra assigned to $CO_2$-$Ar_{17}$. The gap in the observed spectrum is due to a $CO_2$ monomer line (*R*(4) of the ($01^11$)-($01^10$) hot band). The red simulated trace is the sum of the blue, green, and brown ones which have separate *Q*-branches at 2340.472, 2340.465, and 2340.458 $cm^{-1}$, respectively.

The assignments of these bands to $CO_2$-$Ar_{15}$ and $CO_2$-$Ar_{17}$ rely not just on the agreement of rotational constants, but more specifically on the striking intensity alternation of successive lines clearly visible in both bands. This alternation occurs as a direct consequence of the $D_{3h}$ and $D_{5h}$ symmetries of the clusters together with nuclear spin statistics. The PGOPHER software[36] used for the simulated spectra automatically incorporates these spin statistics (given the correct input). The intensity alternation here is somewhat different from that normally encountered, in the sense that we do not observe a simple fixed intensity ratio. First, note that all atoms ($^{12}C$, $^{16}O$, $^{40}Ar$) have zero nuclear spins, which means that only totally symmetric ($A_1'$, $A_1''$) initial ground state rotational levels are allowed. Thus $K''$ can only equal a multiple of 3 for $CO_2$-$Ar_{15}$, or a multiple of 5 for $CO_2$-$Ar_{17}$. Moreover, only even values of $J''$ are allowed for $K''$ = 0, 6, 12, etc. (or 0, 10, 20, etc.), while only odd values of $J''$ are allowed for $K''$ = 3, 9, 15, etc. (or 5, 15, 25, etc.). The $J$-structure is resolved in the spectrum, but not the $K$-structure. The overall result is that lines arising from even $J''$ levels tend to be stronger, especially for lower $J''$ values. But this intensity difference between successive even and odd $J''$ lines diminishes as $J''$ increases, though it never entirely disappears. The washing out of the intensity alternation can be seen in Figs. 3 and 4 for the observed and simulated traces. Overall, the good agreement of observed and simulated spectra clinches the assignments to $CO_2$-$Ar_{15}$ and $CO_2$-$Ar_{17}$.

There is another isomer of $CO_2$-$Ar_{15}$ which could give rise to the observed spectrum of Fig. 3. This isomer has three concentric rings, each consisting of five Ar atoms, giving $D_{5h}$ symmetry. This is just like $CO_2$-$Ar_{17}$, but with the two ‘end’ Ar atoms removed. Its calculated $B$ value is similar to that of the $D_{3h}$ isomer, but its calculated binding energy (5825 $cm^{-1}$) is 281 $cm^{-1}$ less, so it is unlikely to be responsible for the spectrum. We do not think that there are any other isomers of any other cluster size that could explain the spectra of Figs. 3 and 4.

The central region of the $CO_2$-$Ar_{17}$ spectrum (Fig. 4) consists of a strong peak at 2340.472 $cm^{-1}$, which we take as the real $Q$-branch for fitting the spectrum, together with a similar peak at 2340.465 $cm^{-1}$ having about 40% of the intensity of the first, and then another still weaker peak at 2340.457 $cm^{-1}$. Continuing further, the spectrum is obscured by $CO_2$ monomer absorption, leading to a short gap (see Fig. 4), so it is not clear whether the series continues. We attribute this to a vibrational sequence arising from a low-frequency vibrational mode of the $CO_2$-$Ar_{17}$ cluster. The simulation in Fig. 4 (though not the analysis in Table 1) uses this sequence band idea. Based on the intensity of the extra peaks, the low frequency mode would have an energy of the order of 1.3 $cm^{-1}$ if singly degenerate ($A$ symmetry), or 2.4 $cm^{-1}$ if doubly degenerate ($E$). These are very low frequencies, but such a ‘soft’ mode could be possible in a large and relatively weakly-bound cluster like $CO_2$-$Ar_{17}$. Accurate theoretical calculations of such modes are challenging because they involve many intermolecular degrees of freedom. The separation of the $Q$-branch peaks ($\approx$0.007 $cm^{-1}$) happens to be close to $4B$, so their inclusion does not change the simulated spectrum much in the $P$- and $R$-branch regions. If the sequence idea is valid, it implies that each successive excitation of the low frequency mode causes an additional vibrational shift of -0.007 $cm^{-1}$. In the spectrum of $CO_2$-$Ar_{15}$ (Fig. 3), there is only a single $Q$-branch peak. But it appears to have one or more unresolved shoulders on the low frequency side which could also correspond to a similar sequence band progression due to a low frequency mode. Further support for existence of low frequency modes comes from a clear sequence band progression in our spectra for $CO_2$-$Ar_9$ not reported here.

Observed and calculated vibrational frequency shifts, relative to the free $CO_2$ molecule, are listed in Table 2. The present calculations use our (equilibrium) structures together with the differences between the ground state and $v_3 = 1$ potentials (from Cui et al.[23]) summed for each Ar atom surrounding $CO_2$ in a cluster. The calculated shifts from Wang and Xie[7] depend on these same potentials, but used path integral Monte-Carlo (PIMC) methods which should account for zero-point motions, at least in part. The calculated shifts of Severson[3] also used a Monte-Carlo method (QMC) but are of less interest because a realistic difference potential ($v_3 = 1$ minus ground state) was not available at that time. All the calculated shifts agree with experiment in terms of negative sign (red shifts) and approximate magnitude. But interestingly both our calculation and PIMC underestimate the magnitude of the shift for $CO_2$-$Ar_{15}$ and overestimate it for $CO_2$-$Ar_{17}$. This could be related to the significant rearrangement of Ar atoms in going from 15 to 17 together with possible shortcomings in the difference potential given by Cui et al.[23] Vibrational shifts for medium and larger $CO_2$-$Ar_n$ clusters are of obvious interest in exploring the effects of non-additivity, as well as for the interpretation of matrix isolation spectra of $CO_2$ in solid argon, where two lines are observed with characteristic shifts of -4 and -10 $cm^{-1}$.[38] Improved theoretical calculations of vibrational shifts as a function of cluster size and structure will require a highly accurate difference potential suitably integrated over the $CO_2$ intramolecular modes as well as the cluster intermolecular modes.

In spite of continuing theoretical interest, there are almost no previous experimental data on medium-sized clusters of the form $CO_2$-$Ar_n$. Here, rotationally resolved infrared spectra are reported for $CO_2$-$Ar_{15}$ and $CO_2$-$Ar_{17}$, the latter marking completion of the first solvation shell of carbon dioxide in argon. Assignment of the spectra is confirmed by observation of nuclear spin intensity alternation, a direct result of the highly symmetric cage-like structures of these clusters. These structures agree with previous theoretical predictions, and the present results give precise values for the *B* rotational constants and $CO_2$ $v_3$ vibrational frequency shifts which can serve to test future calculations and help to interpret non-additive intermolecular force effects. Apparent *Q*-branch series raise the possibility that these clusters may have low frequency (few $cm^{-1}$) vibrational mode(s), providing an interesting challenge for future theory. Finally, this work opens the possibility of similar studies of CO-$Ar_n$[39] and HCl-$Ar_n$[40] with the present experimental set up.

Table 2. Vibrational shifts and binding energies for $CO_2$-$Ar_{15}$ and $CO_2$-$Ar_{17}$ (in $cm^{-1}$).[a]

| | Vibrational shift | Binding energy |
|---|---|---|
| **$CO_2$-$Ar_{15}$** | | |
| Experiment | -7.163 | |
| Theory (present) | -6.37 | 6106 |
| Theory[6] | | 6088 |
| Theory PIMC[7] | -6.34 | |
| Theory QMC[3] | -6.02 | |
| **$CO_2$-$Ar_{17}$** | | |
| Experiment | -8.671 | |
| Theory (present) | -10.17 | 7063 |
| Theory[6] | | 7033 |
| Theory PIMC[7] | -9.61 | |
| Theory QMC[3] | -6.61 | |

[a] PIMC (path integral Monte Carlo) and QMC (quantum Monte Carlo) shifts were estimated from Fig. 6 of Ref. 7.

**Acknowledgements**

The financial support of the Natural Sciences and Engineering Research Council of Canada is gratefully acknowledged.